\documentstyle[epsf,12pt]{article}
\newcommand{\al}{\alpha'}
\newcommand{\de}{\partial}
\newcommand{\be}{\begin{equation}}
\newcommand{\ba}{\begin{eqnarray}}
\newcommand{\ea}{\end{eqnarray}}
\newcommand{\ee}{\end{equation}}
\newcommand{\db}{\bar{\partial}}

\newcommand{\lr}{\leftrightarrow}
\newcommand{\f}{\frac}
\newcommand{\s}{\sqrt}
\newcommand{\vp}{\varphi}

\newcommand{\ti}{\tilde}
\newcommand{\ap}{\alpha}

\newcommand{\ddd}{\cdot\cdot\cdot}
\newcommand{\no}{\nonumber \\}

\begin{document}
\begin{titlepage}
\thispagestyle{empty}
\begin{flushright}
hep-th/0110099 \\
UT-970 \\
October, 2001 \\
\end{flushright}

\bigskip
\bigskip

\begin{center}
\noindent{\Large \textbf{Orbifolds as Melvin Geometry}}\\
\bigskip
\bigskip
\bigskip
\noindent{
          Tadashi Takayanagi\footnote{
                 E-mail: takayana@hep-th.phys.s.u-tokyo.ac.jp}
                 and Tadaoki Uesugi\footnote{
                 E-mail: uesugi@hep-th.phys.s.u-tokyo.ac.jp} }\\

\bigskip
\it{Department of Physics, Faculty of Science \\ University of Tokyo \\
\medskip
Tokyo 113-0033, Japan}

\end{center}
\bigskip
\begin{abstract}
In this paper we explicitly show that the various noncompact abelian
orbifolds are realized as
special limits of parameters in type II (NSNS) Melvin background and
 its higher dimensional generalizations.
As a result the supersymmetric 
ALE spaces (A-type ${\bf C^2/Z}_N$) and nonsupersymmetric orbifolds in
 type II and type 0 theory are all connected with each other 
by the exactly marginal deformation.
Our results provide new examples of the duality between type II
and type 0 string theory. We also discuss the decay of unstable
 backgrounds in this model which include closed string tachyons.

\end{abstract}
\end{titlepage}

\newpage

\section{Introduction}
\setcounter{equation}{0}
\hspace{5mm}
The Melvin-type magnetic background \cite{Me,GiWi,GiMa,Do,Sa} 
in string theory \cite{BM3,BM,SM,SM2}
has many interesting aspects. This background has the structure
of Kaluza-Klein theory in the sigma model description. 
One of intriguing aspects is that this model in 
superstring gives string theoretic 
examples of supersymmetry breaking which can be exactly solvable.
This model includes three parameters (two magnetic parameters and 
one compactification radius), and in most parts of the parameter space
this non-supersymmetric
model includes closed string tachyons. In particular at the special value
of the magnetic fields with the small radius limit in type II theory
the system
 is equivalent to type 0 string theory (see \cite{Ro,AW,BG,CG,RM,RM2} and 
references
there in). In this way type 0 string theory can be realized as a 
non-supersymmetric
background in type II superstring. 
Recently, this background has also been applied to M-theory
(so called F7-brane \cite{GS,CoHeCo,Em}),
which is properly said as the `9-11' flip of the above mentioned 
NSNS-Melvin background. This leads to the conjectured interpretation of
type 0 theory as the ${\bf S}^1$ compactification of M-theory with the
anti-periodic boundary condition for all fermions \cite{BG}.
These facts may hopefully 
be useful
for the closed string tachyon condensation 
(for recent discussions see e.g. \cite{AS,AS2,Da}).

Another aspect which we would also like to stress in this paper is 
the relation to various noncompact 
orbifolds with a fixed point \cite{orbifold}. 
In particular, by constructing
a higher dimensional generalization of the NSNS-Melvin backgrounds
we find the supersymmetric solvable 
string vacua which include all of (A-type)
ALE orbifolds ${\bf C^2}/{\bf Z}_N$ (see e.g. \cite{ALE,DoMo,BiCrRo2})
as the small radius limits with
various fractional values of the magnetic parameters. We show this fact 
explicitly 
by proving that under such limits the one-loop partition functions
reduce to those of orbifolds. 
For irrational values of 
the magnetic parameters we encounter unfamiliar string vacua which 
preserve half of thirty two supersymmetries and
describe
an `irrationally orbifolded' ten-dimensional non-compact space. 
This model may also be interpreted as a large
$N$ limit of the orbifold ${\bf C^2}/{\bf Z}_N$. The 
higher dimensional Melvin backgrounds we use can also be embedded in 
M-theory by `9-11' flip and these models include the F5-brane \cite{GS,Ur}.

Furthermore, we see that
various two or four dimensional 
non-supersymmetric orbifolds 
in type II and type 0 string
can be also obtained as corners of the moduli space (see Fig.2) of Melvin
geometry in the same way. 
This can be regarded as new examples
of the 
interpolation between the supersymmetric string theories
and the non-supersymmetric ones. For example, one can see that the various
non-supersymmetric orbifolds including the examples in \cite{AS2} can be 
moved into the supersymmetric ones by marginal deformations. 
Such facts are useful for investigating various decay processes in the 
unstable closed string backgrounds as we will see later.

The plan of this paper is as follows. In section two we review some basic 
facts of the Melvin background as a solvable string model. After that, we 
compute its partition function of this model in the small radius limit, and 
see that the result is equivalent to two dimensional orbifolds ${\bf C/Z}_N$
in type II and type 0 string if the magnetic parameters takes rational
values. In section three we generalize the results into higher 
dimensional Melvin backgrounds. In the same limit as before 
we obtain ALE orbifolds and various non-supersymmetric orbifolds. 
In section four we summarize the obtained results and discuss the closed 
string tachyon.

After writing this paper we noticed the preprint \cite{Su2} on the net
which discusses the closed string tachyon in the Melvin background.
We also noted the paper \cite{RuTs}, which has partial overlaps, 
after the present paper appears on the arXiv.

\section{Two Dimensional Orbifolds from Melvin Background}
\setcounter{equation}{0}
\hspace{5mm}
Let us first briefly review the exactly solvable models of type II
superstring in the NS-NS Melvin backgrounds \cite{SM,SM2}. 
See also \cite{BM3,BM} for such models in bosonic string theory.

The target spaces of these models have the structure of Kaluza-Klein theory
and they have the topology $\mbox{M}_3\times{\bf R}^{1,6}$. 
The three dimensional 
manifold $\mbox{M}_3$ is given by ${\bf S}^1$ fibration over ${\bf R}^2$. We 
write the coordinates of ${\bf R}^2$ and ${\bf S}^1$ by $(\rho,\vp)$ 
(polar coordinate) and
$y$ (with radius $R$), respectively.  
This non-trivial fibration is due to two Kaluza-Klein (K.K.)
 gauge fields $A_{\vp}$
and $B_{\vp}$ (see eq.(\ref{KK})), which originate from K.K. reduction
of metric $G_{\vp y}$ and B-field $B_{\vp y}$, respectively. 

The explicit metric and other NSNS fields before the Kaluza-Klein reduction 
are given as follows (we neglect the trivial flat part ${\bf R}^{1,6}$)
\ba
ds^2&=&d\rho ^2+\f{\rho ^2}{(1+\beta^2\rho^2)(1+q^2\rho^2)}d\vp^2+
\f{1+q^2\rho^2}{1+\beta^2\rho^2}(dy+A_{\vp}d\vp)^2, \no 
A_{\vp}&=&\f{q\rho^2}{1+q^2\rho^2},\ \ \ 
B_{\vp y}\equiv B_{\vp}=-\f{\beta\rho^2}{1+\beta^2\rho^2},\ \ \ 
e^{2(\phi-\phi_0)}
=\f{1}{1+\beta^2\rho^2},  \label{KK}
\ea
where $q, \beta$ are the parameters which are proportional to the 
strength of two gauge fields $A_{\vp},~B_{\vp}$ and $\phi_0$ is the 
constant value of the dilaton $\phi$ at $\rho =0$. 

It would be useful to note that if $\beta=0$, then we get the locally flat
metric 
\ba
ds^2=d\rho^2+dy^2+\rho^2(d\vp+qdy)^2.
\ea
However, this background is globally non-trivial because the angle $\vp$ is 
compactified such that its period is $2\pi$. For example, its geodesics lines
$\vp+qy=$constant are spiral and do not return to the same point 
for irrational $qR$ if one goes around the circle ${\bf S}^1$ finite
times. This fact shows the
crucial difference between rational $qR$ and irrational $qR$ 
in physical arguments as we will see. This difference can also be seen 
explicitly in the D-brane spectrum in the NS-NS Melvin backgrounds \cite{TU}.

At first sight the two dimensional sigma model on the above curved 
background (\ref{KK}) for general $q,\beta$ does not seem to be
tractable. However, with appropriate T-duality transformations 
one can solve\footnote{The background eq.(\ref{KK}) satisfies 
the equation of motion even if we take $\al$ corrections into account
\cite{HoTs}.}this sigma model in terms of
free fields \cite{BM,SM}.

In order to see the detailed mass spectrums and the supersymmetry breaking 
in these models let us 
compute the one-loop partition function. This was computed in
the Green-Schwarz formulation \cite{SM,SM2} and we will
transform its expression into that in the NS-R formulation. 

On the world-sheet in the light-cone Green-Schwarz
 formulation, there are eight (real) bosonic fields
$\rho,\vp,Y,X_{i}\ (i=2,3,\ddd,6)$ 
and eight left-moving and right-moving
 fermionic fields $S^a_{L},S^a_{R}\ \ (a=1,2,\ddd,8)$. 
 The fermionic fields
  are divided into 
 two groups $S^{r}_{L,R}$ and $\bar{S}_{L,R}^{r} \ (r=1,2,3,4)$
 according to $U(1)$-charge\footnote{Operators $\hat{J}_{L,R}$ are angular
momentum operators in the $(X'',~\bar{X}'')$
plane.}$\hat{J}_{L,R}=\f{1}{2}$ and $-\f{1}{2}$, 
 where we defined $U(1)$-charge such that $X=\rho e^{i\vp}$ and 
 $\bar{X}=\rho e^{-i\vp}$ also have the charge 
 $\hat{J}_{L}=\hat{J}_{R}=1$ and $-1$ as in \cite{BM,SM2}.
  
Now let us see the calculation of the partition function. 
Introducing the auxiliary vector field $V,\bar{V}$,
 we rewrite the world-sheet action as follows \cite{SM}
\ba
S&=&\f{1}{\pi\al}\int (d\sigma)^2\Bigl[\db\rho\de\rho
+(1+\beta^2\rho^2)V\bar{V}+V(\db Y+\beta\rho^2\db(\vp+qY)+
\f{i\beta}{2}\bar{S}^r_{R}S_{R}^{r})\no
&&\ \ \ -\bar{V}(\de Y-\beta\rho^2\de(\vp+qY)+
\f{i\beta}{2}\bar{S}^r_{L}S_{L}^{r})+\rho^2\de(\vp+qY)\db(\vp+qY)\Bigr]\no
\label{GSi}
&=&\f{1}{\pi\al}\int (d\sigma)^2\Bigl[
(\de+i\beta V+iq\de Y)X(\db-i\beta\bar{V}-iq\db Y)\bar{X}\no
&&+\bar{S}^r_{R}(\de+\f{i\beta}{2}V+\f{iq}{2}\de Y)S_{R}^{r}
+\bar{S}^r_{L}(\db-\f{i\beta}{2}\bar{V}-\f{iq}{2}\db Y)S_{L}^{r}\no
&&\ \ \ +V\bar{V}-\bar{V}\de Y+V\db Y\Bigr]. \label{GS1}
\ea
Here we abbreviate the bosonic parts which come from the trivial
directions ${\bf R}^{1,6}$. Note that if we neglect the fermionic
fields, then we can obtain the (bosonic) sigma model for the Melvin 
background (\ref{KK}) after we integrate out the 
auxiliary field $V,~\bar{V}$\footnote{For the related analysis of the curved 
backgrounds in Green-Schwarz formalism (light-cone gauge) see \cite{FT}.
 In principle it is possible for the world-sheet action in the formalism 
 to include other terms higher than quartic in the fermions. However, these
 specific models are expected to have no higher terms since the free field
 representation is possible as shown by the T-duality. We thank A.A.Tseytlin
 for showing us this observation.}.

Next we would like to 
integrate out the field $Y$.
Then the equation of motion (\ref{GS1}) for $Y$ shows 
$\db V-\de \bar{V}=0$ if we use also the equations of motion for
$X,~\bar{X}$. Therefore we can write $V$ as 
\ba
V=C+\de \ti{Y},\ \ \bar{V}=\bar{C}+\db \ti{Y},
\ea
where $C$ is the constant part ; $\ti{Y}$ is a bosonic field 
which has no zero-modes. Finally we obtain 
(we show only bosonic parts)
\ba
S&=&\f{1}{\pi\al}\int (d\sigma)^2\Bigl[C\bar{C}
-\bar{C}\de Y+C\db Y+ \de\ti{Y} \db\ti{Y}\no
&& +(\de+i\beta C+i\beta\de \ti{Y}+iq\de Y)X\cdot
(\db-i\beta \bar{C}-i\beta\db \ti{Y}-iq\db Y)\bar{X}\Bigr].
\ea
This shows that $X,~S^r_R$ can also be written in terms of free field 
$X'',~S''^r_R$ such that
\ba
X(z,\bar{z})&=&
e^{-i\beta(Cz+\bar{C}\bar{z})-i\beta\ti{Y}-iqY}X''(z,\bar{z}),\no
S^r_R(z,\bar{z})&=&e^{-i\frac{\beta}{2}(Cz+\bar{C}\bar{z})-i\frac{\beta}{2}
\ti{Y}-i\frac{q}{2}Y}S''^r_R(z,\bar{z}).
\label{RF}
\ea
Note also that the terms $\frac{1}{\pi\al}\int(d\sigma)^2(-\bar{C}\de Y+c.c.)$ 
involve only zero-mode parts of
$Y$.
Since the field $Y$ has the periodicity $Y\sim Y+2\pi R$, its zero-mode part 
is quantized in terms of winding numbers $w,w'\in {\bf Z}$ as follows
\ba
Y(\sigma_1,\sigma_2)=\sigma_1 wR+\sigma_2 (w'-w\tau_1)R/\tau_2,
\ea
where $(\sigma_1,\sigma_2)\sim (\sigma_1+2\pi,\sigma_2)\sim 
(\sigma_1+2\pi\tau_1,\sigma_2+2\pi\tau_2)$ 
represents the coordinate of the torus.

Then it is easy to perform the path-integral\footnote{Here 
we have redefined $C,\bar{C}$ such that
$C\to i\bar{C}/\tau_2,\ \ \bar{C}\to -iC/{\tau_2}$.} and the result is
as follows \cite{SM}
\ba
Z(R,q,\beta)&=&(2\pi)^{-7}V_{7}R(\al)^{-5}\int \f{(d\tau)^2}{(\tau_2)^6}
\int (dC)^2 \sum_{w,w'\in {\bf Z}}\ 
\f{|\theta_{1}(\f{\chi}{2}|\tau)|^8}{|\eta(\tau)|^{18}
|\theta_{1}(\chi|\tau)|^2}\no
& &\times \exp\left[-\f{\pi}{\al\tau_2}(
4C\bar{C}-2\bar{C}R(w'-w\tau)+2CR(w'-w\bar{\tau}))\right], \label{PF1}
\ea 
where we have defined 
\ba
\chi=2\beta C+qR(w'-\tau w),\ \ \bar{\chi}=2\beta \bar{C}+qR(w'-\bar{\tau} w).
\ea

The last exponential factor comes from zero modes of $Y$. The theta-function
terms originate from the following path-integral of non-zero modes 
\ba
\f{\det '(\de-\bar{\chi}/(2\tau_2))\det '(\db-\chi/(2\tau_2))}
{\det '(\de)\det '(\db)}&=&\prod_{(n,n')\neq (0,0)}
\f{(n'-\tau n+\chi)(n'-\bar{\tau} n+\bar{\chi})}
{(n'-\tau n)(n'-\bar{\tau} n)}\no
&=&e^{\f{\pi(\chi-\bar{\chi})^2}{2\tau_2}}
\left|\f{\theta_{1}(\chi|\tau)}{\chi\theta_{1}'(0|\tau)}\right|^2.
\ea

In this way we can obtain one-loop partition function in the 
Green-Schwarz formulation. It is easy to check its modular invariance
using theta-function formulas eq.(\ref{TF}).

Now one may ask how to interpret the above result in the NS-R formulation.
If one uses the Jacobi identity eq.(\ref{TF2})
\ba
\theta_{3}(0|\tau)^3\theta_{3}(\chi|\tau)
-\theta_{2}(0|\tau)^3\theta_{2}(\chi|\tau)-
\theta_{4}(0|\tau)^3\theta_{4}(\chi|\tau)=2\theta_{1}(\f{\chi}{2}|\tau)^4
\label{JA1},
\ea
then this explicitly represents the path-integral in the NS-R formulation with
type II GSO-projection. Note that the above partition function does not 
vanish in general and this implies the spacetime supersymmetry breaking. This 
physical result can also be understood in supergravity. To make matters 
simple let us assume $\beta=0$. 
If one goes around the circle ${\bf S}^1$, then 
all of spin $1/2$ fermions receive the phase 
factor $e^{\pm i\pi qR}$ (see eq.(\ref{RF})). Since this
is not equal to one in general, there are no Killing spinors \cite{CG}. 
Only for
$qR\in 2{\bf Z}$ the supersymmetry is preserved and this fact can be easily
seen also from the periodicity (\ref{PR}).

Next we would like to relate the previous result to the mass spectrum in
the operator formulation. Let us assume that the theta-functions are all 
expanded such that each term is an eigen state of the angular momentum 
operators $\hat{J}_{R},\hat{J}_{L}$. 
Here we have defined the charge $\hat{J}_{R},
\hat{J}_{L}$ of the term such that it includes the factor 
$e^{2\pi i\chi \hat{J}_{R}+2\pi i\bar{\chi} \hat{J}_{L}}$ (see eq.(\ref{th})). 
Then by using the Poisson resummation formula
$\sum_{n\in {\bf Z}}\exp(-\pi an^2+2\pi ibn)
=\f{1}{\s{a}}\sum_{m\in {\bf Z}}\exp(-\f{\pi(m-b)^2}{a})$
we can show 
\ba
&&\int (dC)^2\sum_{w,w'\in {\bf Z}}\exp\Biggl[-\f{\pi}{\al\tau_2}(
4C\bar{C}-2\bar{C}R(w'-w\tau)+2CR(w'-w\bar{\tau}))\no
&&\ \ +2\pi i\chi \hat{J}_{R}+2\pi i\bar{\chi} \hat{J}_{L}
+2\pi i\tau \hat{N}_R-2\pi i\bar{\tau} \hat{N}_L\Biggr]\no
&&=\f{\s{(\al\tau_2)^3}}{4R}\sum_{m,w\in {\bf Z}}
\exp(-\pi\al\tau_2M^2+2\pi\tau_1 i(\hat{N}_R-\hat{N}_L-nw)),
\ea
where $M^2$ is the mass spectrum \cite{SM} given by
\ba
\f{\al M^2}{2}&=&\f{\al}{2R^2}(n-qR\hat{J})^2+\f{R^2}{2\al}(w-\f{\al}{R}\beta
\hat{J})^2+\hat{N}_R+\hat{N}_L-\hat{\gamma}(\hat{J}_{R}-\hat{J}_{L}),\no
&&\ \ \ (\hat{\gamma}\equiv\gamma-[\gamma],\ \ \ \gamma\equiv 
qRw+\beta\al(\f{n}{R}-q\hat{J})), \label{MS1}
\ea
with the level matching constraint $\hat{N}_R-\hat{N}_L-nw=0$, where 
$\hat{J}$ is the total angular momentum $\hat{J}=\hat{J}_L+\hat{J}_R$,
and $[\gamma]$ represents the largest integer which is less than $\gamma$.
The integers $\hat{N}_{R,L}$ denote the number of state 
operators which include the zero point energy ($-\frac{1}{2}$ for
NS-sector and $0$ for R-sector).  

Let us see some interesting symmetry of the partition function 
$Z(R,q,\beta)$ \cite{SM}. 
{}From the mass spectrum it is easy to show the T-duality relation
\ba
Z(R,q,\beta)=Z(\f{\al}{R},\beta,q). \label{TD}
\ea 
The interchange of $q$ and $\beta$ corresponds to that of metric
$G_{\vp y}$ and $B_{\vp y}$, which is the essential part of T-duality
transformation in the curved space. Furthermore one can see the
periodicity of $q$ and $\beta$ from eq.(\ref{PF1})
\ba
Z(R,q,\beta)=Z\left(R,~q+2n_1/R,~\beta+2n_2R/\al\right)
~~(n_1,n_2\in{\bf Z}).\label{PR}
\ea

As we have seen, this model does not preserve any supersymmetry
and thus it may be unstable. Therefore we would like to 
know whether the mass spectrum (\ref{MS1}) includes tachyons in the 
NSNS sector.
Then the answer is that it has tachyons if 
neither $qR$ nor $\al\beta/R$ is an integer \cite{SM}. If 
$qR$ or $\al\beta/R$ is an integer, then there is no tachyon for certain
values of the radius \cite{SM}. In particular, for $(qR,\al\beta/R)
\in (2{\bf Z}+1,2{\bf Z})$, 
the model is identical to type IIA(B) theory twisted by
$(-1)^{F_{S}}\cdot\sigma_{1/2}$ \cite{CG} with radius $2R$, 
which is also equivalent to type 0B(A) theory twisted by 
$(-1)^{F_{R}}\cdot\sigma_{1/2}$ \cite{BG} with radius $1/R$.
Here the operators $F_{S},F_{R}$ and $\sigma_{1/2}$ 
represent the spacetime fermion 
number, the world-sheet right-moving fermion number and the half-shift 
operator in the $Y$ direction, respectively. 
If we further take the small radius limit $R\to 0$, we obtain
ten-dimensional type 0 string theory in the T-dualized picture
\cite{BG}. On the other hand, 
if we take the limit $R\to \infty$ with $\beta=0$, 
then the theory is identical
to the ordinary ten dimensional type IIA(B) string theory. See also \cite{Su}
for the similar
relation between supersymmetric and non-supersymmetric 
heterotic string models.

In this way it is interesting to examine the small radius (or
large radius in the T-dualized picture) limit
of the Melvin backgrounds for various values of parameters $q,~\beta$ 
since it is expected to obtain some non-trivial decompactified 
ten-dimensional theories after T-duality. Inspired by this motivation let us
 consider the limit $R\to 0$ and  $\f{\beta\al}{R}\to 0$ with 
the rational value $qR=\f{k}{N}$, 
where $k$ and $N$ are coprime integers. Note that
 this limit is T-dual to $R\to\infty$ and $qR\to 0$ 
with $\f{\beta \al}{R}=\f{k}{N}$
 by using eq.(\ref{TD}). In the former picture with $\beta=0$
  the redefined world-sheet fields (\ref{RF})
 are twisted as follows
\ba
&&X''(\sigma_1+2\pi,\sigma_2)=e^{2\pi i \f{k}{N}w}X''(\sigma_1,\sigma_2),\no
&&S''^r_{R}(\sigma_1+2\pi,\sigma_2)
=e^{\pi i \f{k}{N}w}S''^r_{R}(\sigma_1,\sigma_2),\no
&&Y(\sigma_1+2\pi,\sigma_2)=Y(\sigma_1,\sigma_2)+2\pi Rw. \label{TBC}
\ea
Thus we can see that this string model is equivalent to a ${\bf Z}_{N}$ 
freely acting orbifold\footnote{
For earlier discussions of the related 
Scherk-Schwarz compactification in string 
theory see for example \cite{Ro,AW,KR,KK}.}. 
{}From this we speculate\footnote{In \cite{BM3} the equivalence between 
the Melvin background and 
the orbifold ${\bf C}/{\bf Z}_N$ with H-flux was discussed for other 
case $q=\beta$ with finite radius.}that
the limit $R\to 0$ corresponds to the orbifold ${\bf C}/{\bf Z}_N$.
In order to show this exactly
we investigate the one-loop partition function.
The calculation in the Green-Schwarz formalism is useful
since in this formalism the flip of GSO-projection is automatically
included due to spectral flow\cite{SM}, which is crucial
to determine that the theory is type 0 or type II. 

The partition function (\ref{PF1}) 
 in this limit is given as follows\footnote{If $lk/N$ and $mk/N$ are both
 integers, then $\theta_{1}(\nu_{lm}|\tau)$ does vanish and the partition 
 function will be divergent. This is due to the appearance of the zero
 modes of $(X'',\bar{X}'')\in{\bf R^2}$
 and one should extract this divergence as the volume factor $V_2$.}
  by using the identity (\ref{JA1}) and the 
 quasi-periodicity of theta-functions eq.(\ref{period})
\ba
&&\lim_{R\to 0}Z(R,q,\beta)
=(2\pi)^{-7}V_{7}R(\al)^{-4}\int \f{(d\tau)^2}{4(\tau_2)^5}
\sum_{l,m=0}^{N-1}\sum_{\ap,\beta\in {\bf Z}}\left(\lim_{R\to 0}
e^{-\f{\pi N^2R^2}{\al\tau_2}|\ap-\beta\tau|^2}\right)\no
&&\times\f{|\theta_3(\nu_{lm}|\tau)\theta_{3}(\tau)^3
-(-1)^{k\ap}\theta_2(\nu_{lm}|\tau)\theta_{2}(\tau)^3
-(-1)^{k\beta}\theta_4(\nu_{lm}|\tau)\theta_{4}(\tau)^3|^2}
{4|\eta(\tau)|^{18}
|\theta_{1}(\nu_{lm}|\tau)|^2}, \label{PF2}
\ea
where we have defined $\nu_{lm}=\f{lk}{N}-\f{mk}{N}\tau$, and integers
$l,~m,~\alpha,~\beta$ are given as $w'=N\ap+l,w=N\beta+m$ 
$(l,m=0,1,\ddd,N-1)$. 

If $k$ is an even integer, then the sign factors in front of the 
theta-functions are all plus and we obtain 
\ba
Z(0,q,\beta)&=&V_1V_{7}\int \f{(d\tau)^2}{4(\tau_2)}(4\pi^2\al\tau_2)^{-4}
\no
&&\times\sum_{l,m=0}^{N-1}\f{|\theta_3(\nu_{lm}|\tau)\theta_{3}(\tau)^3
-\theta_2(\nu_{lm}|\tau)\theta_{2}(\tau)^3
-\theta_4(\nu_{lm}|\tau)\theta_{4}(\tau)^3|^2}
{4N|\eta(\tau)|^{18}
|\theta_{1}(\nu_{lm}|\tau)|^2},\label{z1}
\ea
where the divergent factor $V_1$ is also given by 
$V_1=\lim_{R\to 0}\f{2\pi\al}{NR}$ and this
corresponds to the volume
of the noncompact direction. This value of radius $\f{\al}{NR}$ 
is consistent with that 
expected from the boundary condition (\ref{TBC}) by T-duality.
The sums over $l$ and $m$ in this expression 
(\ref{z1}) should be regarded as the ${\bf Z}_N$ projection 
$\frac{1}{N}\sum_{l=0}^{N-1}g^l~~(g=\exp(2\pi i \f{k}{N}\hat{J}))$ and the
sum over twisted sectors, respectively. Therefore the model in this limit is 
identified with the orbifold ${\bf C}/{\bf Z}_N$ 
\cite{c1} in type II string theory. 
As particular examples these orbifolds include those discussed in 
\cite{AS2} ($k=N+1$).

Next let us turn to the case where $k$ is an odd integer. The result is 
\ba
Z(0,q,\beta)&=&V'_1V_{7}\int \f{(d\tau)^2}{4(\tau_2)}(4\pi^2\al\tau_2)^{-4}
\sum_{l,m=0}^{N-1}\no
&&\times\f{|\theta_3(\nu_{lm}|\tau)\theta_{3}(\tau)^3|^2
+|\theta_2(\nu_{lm}|\tau)\theta_{2}(\tau)^3|^2
+|\theta_4(\nu_{lm}|\tau)\theta_{4}(\tau)^3|^2}
{2N|\eta(\tau)|^{18}
|\theta_{1}(\nu_{lm}|\tau)|^2},
\ea
where we have defined\footnote{
The extra factor $1/2$ in comparison with the case of even $k$ is understood
if one notes that
the `GSO-projection' in type 0 is the diagonal ${\bf Z}_2$ projection 
$(1+(-1)^{F_{L}+F_{R}})/2$, while in type II theory
it is give by the ${\bf Z}_2\times{\bf Z}_2$ projection 
$(1+(-1)^{F_{L}})(1+(-1)^{F_{R}})/4$.} $V'_1=V_1/2$.
Thus this model is equivalent to the orbifold ${\bf C}/{\bf Z}_N$ in
type 0 string theory with radius $\f{\al}{2NR}\to \infty$. 
The above results are summarized in Fig.1.

This identification can also be seen from the mass spectrum (\ref{MS1}). 
In the limit of $R\to 0$ we have the constraint $n=\f{k}{N}\hat{J}$. 
This gives the correct ${\bf Z}_N$ orbifold projection. 
The shift of the energy 
$-\hat{\gamma}(\hat{J}_{R}-\hat{J}_{L})
=-(\hat{\f{km}{N}})(\hat{J}_{R}-\hat{J}_{L})$ 
corresponds to the shift of modings in twisted sectors. 
If $k$ is
even, then $\hat{J}$ can be half integer and the NS-R and R-NS sector
are allowed (remember that $\hat{J}$ is the total angular momentum.).
On the other hand if $k$ is odd, then those sectors are not allowed. 
This fact 
gives the difference between type II and type 0 string theory. One can also 
read off the mass of lightest state for each twisted sectors. The result 
is given by for even $k$ (type II)
\ba
\f{\al M^2}{2}=
\left\{
	\begin{array}{cc}
	-\hat{\mu} & \mbox{if}\  \  [\mu] \in \mbox{even} \\
	\hat{\mu}-1  & \mbox{if}\ \  [\mu] \in \mbox{odd}
	\end{array}
\right.  , \label{m2}
\ea
and for odd $k$ (type 0)
\ba
\f{\al M^2}{2}=\min\{\hat{\mu}-1,-\hat{\mu}\},
\ea
where we have defined $\mu=\f{km}{N}$.
The reason that the mass (\ref{m2}) 
depends on whether $[\mu]$ is even or odd is due to the `flip' of
 the sign of GSO projections by the spectral flow \cite{SM}. 
{}From the above results we can conclude that
in the type II orbifolds the tachyon appears in all twisted sectors 
while in type 0 orbifolds 
it does in untwisted sectors as well as in twisted sectors.

Now we would like to return to the detailed 
equivalence between the Melvin background 
with
$qR=\f{k}{N},\beta=0$ for the finite radius 
and the freely acting orbifold (\ref{TBC}). By applying the 
Poisson resummation on $\ap$ to the partition function (\ref{PF2}), 
we can conclude that this special background is equivalent to the ${\bf Z}_N$
orbifold IIA(B)/$\sigma_{\f{1}{N}}\cdot g$ 
 ($g=\exp(2\pi i\frac{k}{N}\hat{J})$) with radius $NR$
(for even $k$) or ${\bf Z}_{2N}$
orbifold IIA(B)/$\sigma_{\f{1}{2N}}\cdot g$ with radius $2NR$ (for odd $k$). 
Here the operators $\sigma_{\f{1}{N}}$ and $\sigma_{\f{1}{2N}}$ mean 
$\frac{1}{N}$ and $\frac{1}{2N}$ shift along ${\bf S}^1$.
The latter case is T-dual to the ${\bf Z}_{N}\times {\bf Z}_{2}$
orbifold 0B(A)/\{$(-1)^{F_R}\cdot\sigma_{\f12},\ 
\tilde{\sigma}_{\f{1}{N}}\cdot g\}$ with radius $\f{1}{NR}$,
where the operator $\tilde{\sigma}_{\f{1}{N}}$ is the T-dual to 
$\sigma_{\f{1}{N}}$. For the special case $N=1$ these results are 
reduced to the results in \cite{BG,CG} if we note the relation 
$g^N=(-1)^{F_{S}}$ for odd k.

Finally let us discuss the limit $R\to 0$ with
irrational values of $qR$. The mass spectrum (\ref{MS1}) shows the
constraint $n-qR\hat{J}=0$ again and this is satisfied iff $n=\hat{J}=0$. 
Remarkably, we can not divide this system into a kind of a 
two dimensional orbifold and one 
dimensional non-compact space. This may also be regarded as a large $N$
limit of ${\bf Z}_N$ orbifold if we are reminded that any irrational number
can be infinitely approximated by rational numbers. Note that taking 
this limit
needs one extra dimension and the non-trivial background is given by the 
non-compact three dimensional manifold. In any case 
we have to say that there are such unfamiliar non-compact string 
backgrounds, which interpolate the previous ${\bf Z}_N$ orbifolds. 

A little analysis of the mass spectrum in this `irrational' model shows
that tachyon fields appear in the sectors $w\neq 0$, and there exist tachyon 
fields whose mass is given by $-1<\f{\al M^2}{2}<-1+\epsilon$ for any 
infinitesimal $\epsilon$.
\begin{figure}[htbp]
 \epsfxsize=110mm
\centerline{\epsfbox{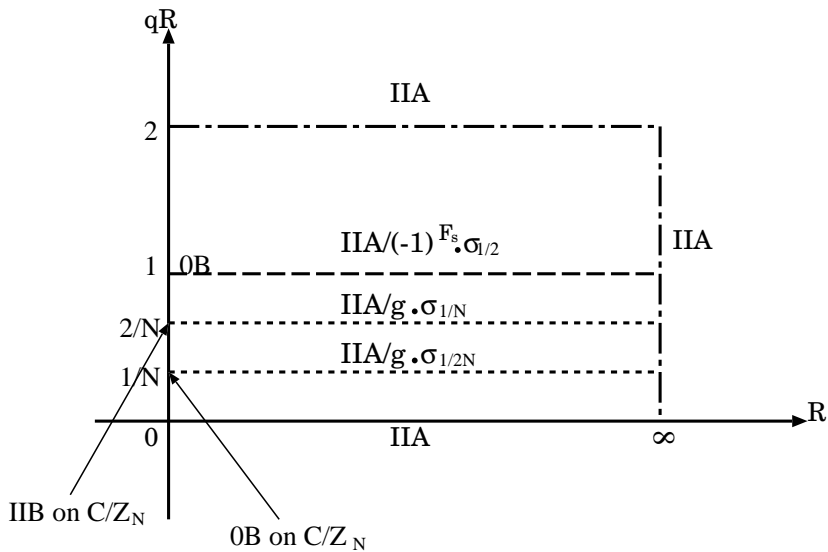}}
 \caption{Moduli space of the string models in type IIA 
             Melvin backgrounds with $\beta=0$.}
  \label{Fig1}
\end{figure}

\section{ALE space from Higher Dimensional Melvin Backgrounds}
\setcounter{equation}{0}
\hspace{5mm}
The closed string backgrounds we have discussed above 
do not preserve any supersymmetry in general. 
However, it is possible to realize
the supersymmetry in higher dimensional generalizations of the Melvin 
background as we will see below.

Let us consider a background of the form
$\mbox{\bf M}_5\times {\bf R}^{1,4}$, where $\mbox{\bf M}_5$ 
is a 
 fibration of ${\bf S}^1\ni Y$ 
 over ${\bf R}^2\times {\bf R}^2\ni (X^1,\bar{X}^1)\times(X^2,\bar{X}^2)$.
At the sigma model level this is possible if one assumes the
higher dimensional generalizations of (\ref{GS1}):
\ba
S&=&\f{1}{\pi\al}\int (d\sigma)^2\Bigl[
(\de+i\beta_1 V+iq_1\de Y)X^1(\db-i\beta_1\bar{V}-iq_1\db Y)\bar{X}^1\no
&&+(\de+i\beta_2 V+iq_2\de Y)X^2(\db-i\beta_2\bar{V}-iq_2\db Y)\bar{X}^2
+V\bar{V}-\bar{V}\de Y+V\db Y\Bigr], \label{GSH}
\ea
where we have omitted the fermion terms.
The explicit metric of this model is given in the appendix B. The special
case $q_1=q_2, \beta_1=\beta_2=0$ can be regarded as the $9-11$ flip of the
supersymmetric F5-brane \cite{GS,Ur}. The free field
representation is again possible almost 
in the same way as before. If we turn to
 the Green-Schwarz formulation, the four of eight (light-cone gauge) 
 spinor fields do not 
 suffer from the phase factor when $\sigma_1$ is shifted by $2\pi$
  if $q_1=q_2,\beta_1=\beta_2$ or $q_1=-q_2,\beta_1=-\beta_2$. 
  Therefore
 we can conclude that in these cases 
 half of thirty two supersymmetries are preserved\footnote{The
mechanism of preserving supersymmetry in our model is
 closely related to the compactified model discussed in \cite{KK}.}.
 From the supergravity viewpoint, we can see this as follows. For 
 simplicity, let us set $\beta_1=\beta_2=0$. Then if we go around the circle
 ${\bf S}^1$, the spinor fields obtain the phase 
$e^{i\pi(\pm q_1\pm q_2)R}$.
 Thus if $q_1=q_2$ or $q_1=-q_2$, there are sixteen Killing spinors. 
 We would also like to mention that 
 similar arguments can also be generalized into seven or nine dimensional
 background $\mbox{\bf M}_7$ and $\mbox{\bf M}_9$ which are fibrations of
$S^1\ni Y$ over ${\bf R}^2\times {\bf R}^2\times {\bf R}^2$ and 
${\bf R}^2\times {\bf R}^2\times {\bf R}^2\times {\bf R}^2$. The `9-11' flip
of these will include the supersymmetric F3-brane, F1-brane 
(see also \cite{Ur,RuTs}).

Another proof of the existence of the supersymmetry is to check the
vanishing of the partition
function, which is equivalent to the Bose-Fermi degeneracy. In the 
path-integral formulation of
Green-Schwarz string one can compute this as before.
The result is given by
\ba
Z(R,q_1,q_2,\beta_1,\beta_2)&=&
(2\pi)^{-5}V_{5}R(\al)^{-4}\int \f{(d\tau)^2}{(\tau_2)^5}
\int (dC)^2 \sum_{w,w'\in {\bf Z}}
\f{|\theta_{1}(\f{\chi_1+\chi_2}{2}|\tau)
\theta_{1}(\f{\chi_1-\chi_2}{2}|\tau)|^4}
{|\eta(\tau)|^{12}
|\theta_{1}(\chi_1|\tau)\theta_{1}(\chi_2|\tau)|^2}\no
&&\times\exp\left[-\f{\pi}{\al\tau_2}(
4C\bar{C}-2\bar{C}R(w'-w\tau)+2CR(w'-w\bar{\tau}))\right],  \label{pfh}
\ea
where we have defined 
\ba
\chi_1=2\beta_1 C+q_1 R(w'-\tau w),\ \ \ \ 
\chi_2=2\beta_2 C+q_2 R(w'-\tau w).
\ea
Thus it is easy to see the vanishing of 
$Z(R,q_1,q_2,\beta_1,\beta_2)$ if $\chi_1=\chi_2$ or $\chi_1=-\chi_2$,
which is equivalent to $q_1=q_2,\beta_1=\beta_2$ or 
$q_1=-q_2,\beta_1=-\beta_2$.

If one wants the partition function in NS-R formalism, then one has only to
note the Jacobi identity eq.(\ref{TF2}) again
\ba
&&2\theta_1\left(\chi_1/2+\chi_2/2|\tau\right)^2
\theta_1\left(\chi_1/2-\chi_2/2|\tau\right)^2\\
&&\ \ =\theta_3(\chi_1|\tau)\theta_3(\chi_2|\tau)\theta_3(0|\tau)^2
-\theta_2(\chi_1|\tau)\theta_2(\chi_2|\tau)\theta_2(0|\tau)^2
-\theta_4(\chi_1|\tau)\theta_4(\chi_2|\tau)\theta_4(0|\tau)^2 \nonumber.
\ea

Next we can obtain the mass spectrum from (\ref{pfh}) by the Poisson 
resummation formula as follows
\ba
\f{\al M^2}{2}&=&\f{\al}{2R^2}(n-q_1 R\hat{J}_1-q_2 R\hat{J}_2)^2
+\f{R^2}{2\al}(w-\f{\al}{R}\beta_1\hat{J}_1-\f{\al}{R}\beta_2\hat{J}_2)^2
\no
&&+\hat{N}_R+\hat{N}_L-\sum_{i=1}^{2}\hat{\gamma_i}
(\hat{J}_{Ri}-\hat{J}_{Li}),
\no
&&\ \ \ \ \ (\hat{\gamma_i}\equiv\gamma_i-[\gamma_i],\ \ \ \gamma_i\equiv
q_i Rw+\beta_i\al(\f{n}{R}-q_1 \hat{J}_1 -q_2 \hat{J}_2),) \label{gam}
\ea
with the level matching constraint $\hat{N}_R-\hat{N}_L-nw=0$. The 
U(1) charges $\hat{J}_1$ and $\hat{J}_2$ are angular momentum operators 
for $X^1$ and $X^2$ directions, respectively.
 
{}From the above expression we can find the T-duality symmetry
$R\lr \frac{\al}{R}$ and $q_i\lr \beta_i$ if $q_1\beta_2=q_2\beta_1$. 
The supersymmetric model
satisfies this condition. We can also prove the periodicity like 
eq.(\ref{PR}).

In the supersymmetric case we have found 
the Bose-Fermi degeneracy.
Therefore this system should have no tachyons. Let us show this explicitly
using the mass spectrum (\ref{gam}). We can assume 
$0<\gamma_{1}=\gamma_{2}< 1$ without any
loss of generality since there are no flip of GSO-projection if 
$\gamma_1=\gamma_2$. Taking the 
GSO-projection into consideration, we obtain the
inequalities $\hat{J}_{1R}+\hat{J}_{2R}\leq \hat{N}_{R}$
and $-\hat{N}_{L}\leq \hat{J}_{1L}+\hat{J}_{2L}$. 
Then it is easy to 
see  
in the NSNS sector
\ba
\f{\al M^2}{2}=(\hat{N}_R-\gamma \hat{J}_{1R}-\gamma \hat{J}_{2R})+
(\hat{N}_L+\gamma_1 \hat{J}_{1L}+\gamma_2 \hat{J}_{2L})+\f{\al}{4}
(P_{R}^2+P_{L}^2)\ \geq\  0.
\ea

Now we would like to discuss the relation between the above models and 
orbifolds. We consider both supersymmetric and non-supersymmetric cases.
Let us take the limit $R\to 0$ with $\f{\beta_1\al}{R}\to 0$ and 
$\f{\beta_2\al}{R}\to 0$, and assume that $q_1R$ and $q_2R$ are
fractional. 
We can write\footnote{Here we assume that there is no positive 
integer other than one which divides all of the three integers $N,k_1$
and $k_2$.}
them as $q_1R=\f{k_1}{N}$ and $q_2R=\f{k_2}{N}$. Then the partition
function in this limit becomes as in the previous calculations
\ba
&&\lim_{R\to 0} Z(R,q_1,q_2,\beta_1,\beta_2)=
(2\pi)^{-5}V_{5}R(\al)^{-3}\int \f{(d\tau)^2}{16(\tau_2)^4}
\left(\lim_{R\to 0} \sum_{\ap,\beta \in {\bf Z}}
e^{-\f{\pi N^2R^2}{\al\tau_2}|\ap-\beta\tau|^2}\right)
|\eta(\tau)|^{-12}
\no
&&\hspace{3cm}\times \sum_{l,m=0}^{N-1} \Bigl|\theta_3(\nu^1_{l,m}|\tau)
\theta_3(\nu^2_{l,m}|\tau)\theta_3(\tau)^2
-(-1)^{(k_1+k_2)\ap}
\theta_2(\nu^1_{l,m}|\tau)\theta_2(\nu^2_{l,m}|\tau)\theta_2(\tau)^2 \no
&&\hspace{4cm} -(-1)^{(k_1+k_2)\beta}
\theta_4(\nu^1_{l,m}|\tau)\theta_4(\nu^2_{l,m}|\tau)\theta_4(\tau)^2
\Bigr|^2
\cdot\left|\theta_{1}(\nu^1_{l,m}|\tau)\theta_{1}
(\nu^2_{l,m}|\tau)\right| ^{-2}
,\no
\ea
where we have defined $\nu^1_{l,m}=\f{k_1}{N}(l-m\tau)$ and 
$\nu^2_{l,m}=\f{k_2}{N}(l-m\tau)$.

If $k_1+k_2$ is even, then we get the result
\ba
&&Z(0,q_1,q_2,\beta_1,\beta_2)=V_1V_5
\int \f{(d\tau)^2}{4\tau_2}
(4\pi^2\al\tau_2)^{-3}
\sum_{l,m=0}^{N-1}\no
&&\times \f{|\theta_3(\nu^1_{l,m}|\tau)
\theta_3(\nu^2_{l,m}|\tau)\theta_3(\tau)^2
-\theta_2(\nu^1_{l,m}|\tau)\theta_2(\nu^2_{l,m}|\tau)\theta_2(\tau)^2 
-\theta_4(\nu^1_{l,m}|\tau)\theta_4(\nu^2_{l,m}|\tau)\theta_4(\tau)^2|^2}
{4N\ |\eta(\tau)|^{12}
|\theta_{1}(\nu^1_{l,m}|\tau)\theta_{1}
(\nu^2_{l,m}|\tau)|^{2}}.\no \label{PF22}
\ea
Thus we have the abelian non-compact 
four dimensional orbifolds 
${\bf C}^2/{\bf Z}_N$ in type II string theory. These include both 
the supersymmetric and non-supersymmetric orbifolds. The former correspond
to the values $k_1=\pm k_2$ and it is easy to see that for fixed $N$ 
the partition functions (\ref{PF22}) for each $k_1,k_2$ 
give the same value. This represents the $A_{N-1}$-type 
ALE space (for a review see \cite{ALE}) in the orbifold limit.

The other orbifolds are all non-supersymmetric and the tachyon can appear
only in the twisted sectors. These include the examples  
discussed in \cite{AS2}, where the specific non-supersymmetric orbifolds
are argued to decay into ALE spaces. Our results show that both such 
non-supersymmetric orbifolds and supersymmetric ALE orbifolds (including 
the type II string in flat space $N=1$) are connected
in the moduli space of solvable superstring models if we compactify one
direction. In other words, these two different kind of orbifolds are all 
obtained from each other by marginal deformations. We will discuss the
decay of such nonsupersymmetric orbifolds in the last section. 

Next we turn to the case where $k_1+k_2$ is odd. The partition function 
is given by
\ba
&&Z(0,q_1,q_2,\beta_1,\beta_2)=V'_1V_7
\int \f{(d\tau)^2}{4\tau_2}
(4\pi^2\al\tau_2)^{-3}
\sum_{l,m=0}^{N-1}\no
&&\times \f{|\theta_3(\nu^1_{l,m}|\tau)
\theta_3(\nu^2_{l,m}|\tau)\theta_3(\tau)^2|^2
+|\theta_2(\nu^1_{l,m}|\tau)\theta_2(\nu^2_{l,m}|\tau)\theta_2(\tau)^2|^2 
+|\theta_4(\nu^1_{l,m}|\tau)\theta_4(\nu^2_{l,m}|\tau)\theta_4(\tau)^2|^2}
{2N|\eta(\tau)|^{12}
|\theta_{1}(\nu^1_{l,m}|\tau)\theta_{1}
(\nu^2_{l,m}|\tau)|^{2}}.\no \label{PF23}
\ea

This explicitly shows that the systems now considered are equivalent to the
non-compact 
four dimensional orbifolds ${\bf C}^2/{\bf Z}_N$ in type 0 string 
theory\footnote{Such orbifolds are considered in the context of D-branes
in \cite{BCR}.}.
The above result shows that orbifolds in type 0 theory are connected to
those in type II theory. In particular, this shows that 
various orbifolds in type 0 theory
can be regarded as non-supersymmetric backgrounds in type II string theory.

Then let us discuss the tachyons in these orbifolds.
The mass of the lightest state 
is given by for even $k_1+k_2$ (type II)
\ba
\f{\al M^2}{2}=
\left\{
	\begin{array}{cc}
 -|\hat{\mu_1}-\hat{\mu_2}| & 
 \mbox{if}\ \ ([\mu_1],[\mu_2]) \in \mbox{(even,even) or (odd,odd)} \\
  \hat{\mu_1}+\hat{\mu_2}-1 & 
  \mbox{if}\ \  ([\mu_1],[\mu_2]) \in \mbox{(even,odd) or (odd,even)}\\
 \end{array}
\right.  , \label{m4}
\ea
and for odd $k_1+k_2$ (type 0)
\ba
\f{\al M^2}{2}=\min\{\hat{\mu_1}+\hat{\mu_2}-1,-|\hat{\mu_1}-\hat{\mu_2}|\},
\ea
where we have defined $\mu_1=\f{k_1m}{N}$ and $\mu_2=\f{k_2m}{N}$.
 These results show that in both orbifolds some of the twisted sectors will
 contain tachyon and in type 0 orbifolds tachyon also appears in
 the untwisted sector. The type II string theory 
on ALE orbifolds $k_1=\pm k_2$ are only examples of tachyon less orbifolds.

Finally, if we turn to the small radius limit with irrational 
values of $q_1R$ and $q_2R$, we obtain the `irrationally orbifolded'
 noncompact space (or a kind of a `large $N$ limit of 
 the orbifolds ${\bf C}^2/{\bf Z}_{\infty}$') 
as in the original Melvin background. For 
specific values $q_1=\pm q_2$ this
 background preserves the half of thirty two supersymmetries and 
is connected to ALE spaces. 
 
It will be also interesting to consider the brane picture of
the general supersymmetric backgrounds with $\beta_{1,2}\neq 0$, 
which includes the previous orbifolds as special examples by T-duality.
These models have the non-trivial $H$-flux and 
dilaton gradient. We left its relation to NS5-branes as a future problem.

\begin{figure}[htbp]
     \epsfxsize=90mm
  \centerline{\epsfbox{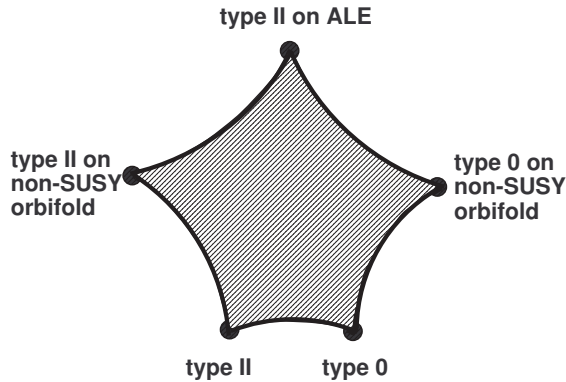}}
	\caption{Moduli space of the string models in the higher dimensional
Melvin background.}
	\label{Fig2}
\end{figure}

In this way we have shown that the various orbifolds both 
non-supersymmetric and supersymmetric
 are included in the moduli space of the higher dimensional 
 Melvin background (see Fig.\ref{Fig2}).

\section{Conclusions and Discussions}
\setcounter{equation}{0}
\hspace{5mm}
In this paper we have discussed the conformal field theory of the 
Melvin background and its higher dimensional generalization. 
In the latter one we found new examples of solvable supersymmetric
string models on 
non-compact five dimensional manifolds (see eq.(\ref{ME})) with H-flux. 
We have shown that various abelian orbifolds in type II
and type 0 string theory can be obtained in the small radius (or
large radius in the T-dualized picture) limit of 
the Melvin background with the rational values of the magnetic parameters.
Even though the examples we have examined are two and four 
dimensional orbifolds 
${\bf C}/{\bf Z}_N,\ {\bf C}^2/{\bf Z}_N$, our results will be
 easily generalized
into much higher dimensional orbifolds 
${\bf C}^{n}/{\bf Z}_N,\  n=3,4$. 
Other abelian orbifolds such as 
${\bf C}^{n}/{\bf Z}_{N_{1}}\times {\bf Z}_{N_{2}}
\times\ddd\times {\bf Z}_{N_{K}}$ 
can also be obtained if
we replace ${\bf S}^1$ with $K$ dimensional tori.
Non-abelian orbifolds (e.g. D,E type) will deserve future study. 
The T-dual counterparts of these examples are also interesting 
since the presence of the H-flux is essential in such models.
They seem to have some relevance to NS5-branes. 

One of the important lessons which can be obtained from the above 
arguments 
is that we can view both non-supersymmetric and supersymmetric orbifolds
as marginal 
deformations of flat type II or type 0 string theory. 
This can be seen as a new example of the duality 
relation between stable supersymmetric string theories 
and unstable string theories which include tachyons. Thus it is intriguing 
to explore the application to the decay of unstable spacetime
or closed string tachyon condensation.

Let us give an very intuitive idea about the decay processes of 
these unstable backgrounds. 
Since we discuss only 
exactly marginal deformations, we have no tree level potential. Instead
the leading quantity which determines the stability will be
the one-loop amplitude (or cosmological constant). Even though this
value is IR divergent in the general parameter region of the Melvin
background, this divergence
 can be estimated by the tachyon mass whose absolute value is largest. 
Thus we can expect
 that the unstable tachyonic background will decay into another background
 which is less tachyonic (the absolute value of the tachyon mass
square is smaller) and eventually into the tachyon less background.
Let us take the example of the orbifold ${\bf C}/{\bf Z}_N$. If we are
reminded of the moduli space (see Fig.1 or Fig.2) of 
Melvin background, we find many decay routes.
The most straightforward way is to decay into the ordinary type II string 
vacuum by the spontaneous (de)compactification of $Y$ direction.
Another possibility is the shift of the magnetic parameter. However 
the infinitesimal shift will make the value of $qR$ irrational. Since such a 
background is more tachyonic as we saw in the last of section two, this decay
route is not preferable. On the other hand, the decay
mode from ${\bf C}/{\bf Z}_{2l+1}$ into 
${\bf C}/{\bf Z}_{2l-1}$ discussed in \cite{AS2} 
is more favorable because the absolute value of tachyon mass decreases. 
Note that the above arguments are not involved in the
 real closed string tachyon condensation by relevant perturbations as 
contrasted with \cite{AS2}. This viewpoint may be supported by the
observed absence of tree level tachyon potential \cite{BSFT,RM}. 

Another important lesson is that the limits of the Melvin backgrounds 
depend very sensitively on whether the value of the magnetic flux is
rational or irrational. In the latter case one can find unfamiliar 
supersymmetric string backgrounds, 
which can be regarded as a `large $N$ orbifold'. The 
existence of two remarkably different kinds of limits can also be seen
in the D-brane spectrums on such backgrounds as we will discuss in 
our forthcoming paper \cite{TU}.

\bigskip
\begin{center}
\noindent{\large \bf Acknowledgments}
\end{center}
We are grateful to S. Yamaguchi for useful comments and S. Kawamoto 
for showing us some interesting related topics.
We also thank A.A.Tseytlin for e-mail correspondence.
T.T. is supported by JSPS Research Fellowships for Young Scientists.

\appendix
\setcounter{equation}{0}

\section{Identities of Theta Functions }

Here we summarize the formulae of $\theta$-functions. First define the 
following $\theta$-functions:
\ba
\eta(\tau)&=&q^{\f{1}{24}}\prod_{n=1}^{\infty}(1-q^n),\no
\theta_{1}(\nu,\tau)&=&2q^{\f18}\sin(\pi\nu)\prod_{n=1}^{\infty}(1-q^n)
(1-e^{2i\pi\nu}q^{n})(1-e^{-2i\pi\nu}q^{n}),\no
\theta_{2}(\nu,\tau)&=&2q^{\f18}\cos(\pi\nu)\prod_{n=1}^{\infty}(1-q^n)
(1+e^{2i\pi\nu}q^{n})(1+e^{-2i\pi\nu}q^{n}),\no
\theta_{3}(\nu,\tau)&=&\prod_{n=1}^{\infty}(1-q^n)
(1+e^{2i\pi\nu}q^{n-\f12})(1+e^{-2i\pi\nu}q^{n-\f12}),\no
\theta_{4}(\nu,\tau)&=&\prod_{n=1}^{\infty}(1-q^n)
(1-e^{2i\pi\nu}q^{n-\f12})(1-e^{-2i\pi\nu}q^{n-\f12}), \label{th}
\ea
where we have defined $q=e^{2i\pi\tau}$.

Next we show the modular properties as follows
\ba
\eta(\tau)&=&(-i\tau)^{-\f12}\eta(-\f{1}{\tau}),\ \ \theta_{1}(\nu,\tau)
=i(-i\tau)^{-\f12}e^{-\pi i\f{\nu^2}{\tau}}\theta_{1}
(\nu/\tau,-\f{1}{\tau}), \no
\theta_{2}(\nu,\tau)&=&(-i\tau)^{-\f12}e^{-\pi i\f{\nu^2}{\tau}}
\theta_{4}(\nu/\tau,-\f{1}{\tau}), \ \ \theta_{3}(\nu,\tau)
=(-i\tau)^{-\f12}e^{-\pi i\f{\nu^2}{\tau}}\theta_{3}(\nu/\tau,-\f{1}{\tau})
, \no\theta_{4}(\nu,\tau)&=&(-i\tau)^{-\f12}e^{-\pi i\f{\nu^2}{\tau}}
\theta_{2}(\nu/\tau,-\f{1}{\tau}). \label{TF}
\ea

Their quasi periodicity is also given by
\ba
\label{period}
\theta_1(\nu+\tau|\tau)&=&-e^{-2\pi i\nu-\pi i \tau}\theta_1(\nu|\tau), \no
\theta_2(\nu+\tau|\tau)&=&e^{-2\pi i\nu-\pi i \tau}\theta_2(\nu|\tau), \no
\theta_3(\nu+\tau|\tau)&=&e^{-2\pi i\nu-\pi i \tau}\theta_3(\nu|\tau), \no
\theta_4(\nu+\tau|\tau)&=&-e^{-2\pi i\nu-\pi i \tau}\theta_4(\nu|\tau). 
\ea

It is also useful to note the Jacobi's identity
\ba
\prod_{a=1}^{4}\theta_{3}(\nu_a|\tau)-\prod_{a=1}^{4}\theta_{2}(\nu_a|\tau)
-\prod_{a=1}^{4}\theta_{4}(\nu_a|\tau)+\prod_{a=1}^{4}\theta_{1}(\nu_a|\tau)
=2\prod_{a=1}^{4}\theta_{1}(\nu'_a|\tau), \label{TF2}
\ea
where we have defined
\ba
&&2\nu'_1=\nu_1+\nu_2+\nu_3+\nu_4,\ \ 2\nu'_2=\nu_1+\nu_2-\nu_3-\nu_4,\no \ \ 
&&2\nu'_3=\nu_1-\nu_2+\nu_3-\nu_4, \ \ 2\nu'_4=\nu_1-\nu_2-\nu_3+\nu_4.
\ea

\section{Explicit Metric of the Higher Dimensional Model}
\setcounter{equation}{0}
Let us define the polar coordinates as $X^1=\rho e^{i\vp},\ \ X^2=
re^{i\theta}$.
The metric (in string frame) is given by 
\ba
(ds)^2&=&d\rho^2+dr^2+\f{1+\beta_2 r^2}{F(r,\rho)}\rho^2 d\vp^2
-2\f{\beta_1\beta_2 r^2\rho^2}{F(r,\rho)}d\vp d\theta
+\f{1+\beta_1^2 \rho^2}{F(r,\rho)}
r^2 d\theta^2 \no
&&+
\f{G(r,\rho)}{F(r,\rho)}
dy^2+2\f{q_1\rho^2(1+\beta_2^2 r^2)-\beta_1\beta_2 q_2\rho^2 r^2}{F(r,\rho)}
d\vp dy\no
&&+2\f{q_2 r^2(1+\beta_1^2 \rho^2)-\beta_1\beta_2 q_1\rho^2 r^2}
{F(r,\rho)}d\theta dy,\no
&=&\f{G(r,\rho)}{F(r,\rho)}
(dy+A_{\vp}d\vp+A_{\theta}d\theta)^2-\f{2r^2\rho^2}{G(r,\rho)F(r,\rho)}
\left(\beta_1\beta_2+q_1q_2 F(r,\rho)\right)d\vp d\theta \no
&&+d\rho^2+\f{\rho^2 }{G(r,\rho)F(r,\rho)}\left(1+\beta_2^2 r^2+q_2^2 r^2
+q_2^2 \beta_2^2 r^4+\beta_1^2 q_2^2 \rho^2 r^2 \right)d\vp ^2 \no
&&+dr^2+\f{r^2}{G(r,\rho)F(r,\rho)}\left
(1+\beta_1^2\rho^2+q_1^2\rho^2+q_1^2\beta_1^2
\rho^4+\beta_2^2q_1^2\rho^2 r^2 \right)d\theta^2, \label{ME}
\ea
where we have defined 
\ba
F(r,\rho)\equiv 1+\beta_1^2\rho^2+\beta_2^2 r^2,\ \  \
G(r,\rho)\equiv 
1+(\beta_1 q_2-\beta_2 q_1)^2\rho ^2 r^2 +q_1^2\rho^2 +q_2^2r^2, \no
A_{\vp}=\f{q_1\rho^2+\beta_2(q_1\beta_2-q_2\beta_1)\rho^2 r^2}
{G(r,\rho)},\ \ \ 
A_{\theta}=\f{q_2 r^2+\beta_1(q_2\beta_1-q_1\beta_2)\rho^2 r^2}{G(r,\rho)}.
\ea
The B-field and the dilaton is 
\ba
B_{\vp}=-\f{\beta_1 \rho^2}{F(r,\rho)},\ \ 
B_{\theta}=-\f{\beta_2 r^2}{F(r,\rho)},\ \ 
e^{2(\phi-\phi_0)}=\f{1}{F(r,\rho)}.
\ea
One can check that the curvature of the above metric is not singular.

It is not so difficult to show that these satisfy the equations of
motion in supergravity
\ba
&&R_{\mu\nu}+\nabla_{\mu}\nabla_{\nu}\phi-\f{1}{4}
H_{\mu\ap\beta}H_{\nu}^{\ap\beta}=0,\no
&&-\f{1}{2}\nabla^2\phi+\nabla_{\mu}\phi\nabla^{\mu}\phi-\f{1}{24}
H_{\mu\nu\rho}H^{\mu\nu\rho}=0,\no
&&
\nabla_{\rho}H^{\rho}_{\mu\nu}-2(\nabla_{\rho}\phi) H^\rho_{\mu\nu}=0.
\ea

It is also worth noting that we can relate 
$B_{\vp,\theta}$ to $A_{\vp,\theta}$ by T-duality: $R\lr\frac{\al}{R}$
and $q_i\lr\beta_{i}$
if $(\beta_1 q_2-\beta_2 q_1)=0$. This includes the supersymmetric cases
$q_1=\pm q_2,\ \ \beta_{1}=\pm\beta_{2}$.

\end{document}